\documentclass[11pt,a4paper]{article}
\usepackage{epsfig}
\begin{document}

\begin{titlepage}

\hbox{}\vspace{1in}
\begin{center}

\textsc{ \Large Pulse Sequences for NMR Quantum Computers:\\ How to Manipulate 
Nuclear Spins
          While Freezing the Motion of Coupled Neighbours}

\vspace{1in}

\textbf{Noah Linden$^{a,b,}$\footnote{\tt nl101@newton.cam.ac.uk},
Herv\'e Barjat$^{c,}$\footnote{\tt hb232@cus.cam.ac.uk},
Rodrigo J. Carbajo$^c$ and
 Ray Freeman$^{c,}$\footnote{{\tt rf110@cus.cam.ac.uk}; {\rm author to whom
correspondence should be addressed.}}}

\vskip 1truecm

$^a$Isaac Newton Institute for Mathematical
Sciences, 20 Clarkson Road, Cambridge, CB3 0EH, UK\\
$^b$Department of Applied Mathematics and Theoretical Physics, Silver Street,
Cambridge CB3 9EW, UK\\
$^c$Department of Chemistry, Lensfield Rd,
Cambridge CB2 1EW, UK.

\vspace{1in}

\bigskip


\end{center}

\begin{center} \textsc{\large Abstract} \end{center} \medskip
We show how to divide a coupled multi-spin system into an
small subset of ``active'' spins that evolve under chemical shift or scalar
coupling operators, and a larger subset of ``spectator'' spins which are
returned to their initial states, as if their motion had been temporarily 
frozen.  This allows us to implement basic one-qubit and 
two-qubit operations from which general operations on $N$-qubits can be 
constructed, suitable for quantum computation.
The principles are illustrated 
by experiments on the three coupled protons of 2,3-dibromopropanoic acid, 
but the method 
is applicable to any spin-1/2 nuclei and to systems containing arbitrary 
numbers of 
coupled spins.

\end{titlepage}

{\openup 5pt

\section{Introduction}

In this letter we address the problem of how to use NMR techniques to 
manipulate subsets of spins in a multi-spin system while leaving the states 
of all the other coupled spins unchanged.  We have in mind the use of these 
manipulations to generate pulse sequence modules which will be useful for 
NMR quantum computation 
\cite{Cory97a,Cory98,Gershenfeld97,Chuang97,Chuang98a,Chuang98b,%
Jones98a,Jones98b,Linden98a}.  This leads to an important difference of 
emphasis 
from conventional NMR spectroscopy where one 
can assume that the states of the spins before the manipulation are known, and 
indeed one often arranges these states (for example as $z$-magnetization) so 
that the manipulation has a simple 
effect on the spins.  
One is certainly not concerned about the effect that the pulse sequence might 
have  on spins which were prepared in different states. Furthermore, one is 
typically only interested in the 
effect of the manipulation on a subset of the spins and it is irrelevant how 
the 
other spins evolve.
In contrast, in applications to NMR quantum computing, the state of the system 
before the given pulse 
sequence module is assumed unknown.  The module is a logic 
gate which is designed to carry out a specified unitary 
transformation which can be applied to any input, the remaining qubits must 
remain passive spectators.

A system of
$N$ spins  evolves under a Hamiltonian 
which includes  chemical shifts for each spin and scalar coupling terms for 
every pair of spins.  
Thus if one wishes to manipulate a particular spin, and the manipulation is not 
effectively instantaneous, all the other spins will evolve, in a complicated, 
coupled, manner.  We will show 
here how to apply patterns of pulses so that only the desired chemical shifts 
and 
scalar couplings affect the evolution of the system {\em
as if we had ``switched off'' all the other terms in the Hamiltonian}.  For 
example, we will show how to produce 
pulses whose net effect is to allow only one spin to precess under its 
chemical shift: no matter what the input states of the other spins are, 
at the end of the pulse sequence they will be returned to their initial states.

For concreteness we will use a weakly-coupled three spin ($ISR$) system to 
demonstrate the 
theoretical and  experimental results.  This system has the ``free'' 
Hamiltonian 
(i.e. the Hamiltonian in the absence of pulses) of the form
\begin{eqnarray}
{\cal H}_0 =  \omega_I I_z+\omega_S S_z+\omega_R R_z + 2\pi J_{IS} I_z S_z
+ 2\pi J_{IR} I_z R_z + 2\pi J_{SR} S_z R_z.
\end{eqnarray}
Using patterns of refocusing pulses  
we will show how to create unitary operations on the spin system  as 
if  many of the terms  in the Hamiltonian were zero.  
  We will demonstrate 
theoretically and experimentally how to generate three basic operations:
\begin{itemize}
\item{\bf O1} Chemical shift evolution of a single spin
\item{\bf O2} Rotation of a single spin by an arbitrary angle about a general 
axis (this generalises {\bf O1})
\item{\bf O3} Evolution under scalar coupling leading to the two-qubit 
operation 
whose product operator representation is
$\exp(i\phi 2I_z S_z)$
\end{itemize}
Straightforward generalisations of the sequences will produce analogous modules 
to 
{\bf 
O1}, {\bf O2} and  {\bf O3} for systems of any number of spins and these 
modules are sufficient to generate any unitary 
operation  on $N$ qubits.
This follows from the important result in 
\cite{Barenco95} that  any arbitrary unitary transformation of 
$N$ spins can be built up from two simple building blocks: these may be 
taken to be, for example, an arbitrary rotation of a single spin, and the 
two-qubit 
operation $\exp(-i{\pi\over 2} 2I_z S_z)$.  The reason for this particular 
choice 
of 
basic 
operations is that they may easily be generated using the 
natural evolution of the system.  We note that the two-qubit  operation 
$\exp(-i{\pi\over 2} 2 I_z S_z)$, which has matrix representation
\begin {eqnarray}
e^{-i{\pi\over 4}}\left(\begin{array}{cccc}
1& 0& 0& 0\\
0&i& 0& 0\\
0& 0&i& 0\\
0& 0& 0&1\\
\end{array}\right),
\end{eqnarray}
 is closely related to the controlled-not (CNOT) operation on two qubits.

The first operation {\bf O1} is implemented by the pulse sequence given in Fig 1.
The sequence of refocussing $\pi$ pulses on spins $R$ and $S$ creates a unitary 
evolution of the system  as if the only
non-zero term in the Hamiltonian were $\omega_I I_z$.  Thus the effect of the
sequence is to act on the complete system with the unitary operator  
$
e^{i\phi I_z}
$.  As an example of how the generalisation to $N$-spins works, consider the 
case 
of four spins labelled $ISRQ$.  Evolution of the $I$ spin alone can be arranged 
during a period of $16\tau$ by having $\pi$ pulses on $S$ at $(8m+4)\tau,\ 
m=0\ldots 1$, on $R$ at $(4m +2)\tau,\ m=0\ldots 3$, and on $Q$ at 
$(2m+1)\tau,\ 
m=0\ldots 7$.

Returning to the $ISR$ system, we see that we can convert the phase evolution 
sequence (a rotation about the 
$z$ axis)  into a 
general excitation pulse {\bf O2}.  The $I$-spin magnetization may be rotated 
through 
an angle $\phi$ about 
a tilted axis in the $xz$ plane by enclosing the unitary operator $e^{i\phi 
I_z}$ between 
two hard pulses applied 
along the $-y$ and $+y$ axes:
\begin{eqnarray}
e^{-i\beta (I_y + S_y + R_y)}\ e^{i\phi I_z}\ e^{+i\beta (I_y + S_y + R_y)};
\label{beta_xy}
\end{eqnarray}
the hard pulses are effectively instantaneous so that no evolution under ${\cal 
H}_0$  takes place.
This tilts the rotation axis away from $+z$ through an angle $\beta$ in the 
$xz$ plane.   The rotation can be made completely general by 
enclosing sequence 
(\ref{beta_xy}) between two $z$-rotation {\bf O1} modules:
\begin{eqnarray}
e^{-i\gamma I_z}\ e^{-i\beta (I_y + S_y + R_y)}\ e^{i\phi I_z}\ e^{+i\beta (I_y 
+ S_y + R_y)}\ e^{+i\gamma I_z}\ .
\label{general_one_spin}
\end{eqnarray}

Rotation now takes place about a tilted radiofrequency field in a vertical 
plane that subtends 
an azimuthal angle $\gamma$ with respect to the $+x$ axis.
As a specific example we will demonstrate experimentally the case $\phi = \pi$, 
$\beta =\pi/4$, $\gamma =0$.

The final basic operation is the two-qubit operation {\bf O3}. The sequence of 
refocussing pulses is given in Fig 2.  We note that this sequence, together 
with the previous ones, can be used to create the unitary operation whose 
total effect is to perform a CNOT operation between $I$ and $S$ 
(with the $R$ spin unaffected).  Recall that the effect of {\bf O3}
is, in terms of product operators, 
$
\exp(i\phi 2I_z S_z)$.
The CNOT operation on two spins may be represented by the 
unitary matrix
\begin {eqnarray}
U_{\rm CNOT}=\left(\begin{array}{cccc}
1& 0& 0& 0\\
0&0& 0& 1\\
0& 0& 1& 0\\
0& 1& 0&0\\
\end{array}\right).
\end{eqnarray}
The corresponding product operator representation is
\begin{eqnarray}
U_{\rm CNOT} 
&=&e^{-i{\pi\over 4}}\ e^{+i{\pi\over 2} I_z}\  e^{+i{\pi\over 2}  I_x } 
e^{-i{\pi\over 2} 2I_x S_z } 
\nonumber\\
&=&e^{-i{\pi\over 4}}\ e^{+i {\pi\over 2}I_z}\  e^{-i{\pi\over 2}  I_y }
e^{+i{\pi\over 2}  
I_z } e^{-i{\pi\over 2} 2I_z 
S_z }e^{+i {\pi\over 2} I_y }.
\end{eqnarray}
It can be seen that each term in the product (except for the irrelevant overall 
phase) can be created using {\bf O1}, {\bf 
O2} or {\bf O3}.

Thus the pulse sequences we describe in this letter give us a set of basic 
modules which are sufficient to produce an arbitrary unitary operation on a 
system of $3$ qubits.   In practice there may be more straightforward 
pulses which suggest themselves as NMR implementations of the essential parts 
of given gates.  However there are often ``corrections'' which need to be made 
to implement the exact unitary operator required \cite{Cory98,Linden98a}.  
Consider for example the controlled-controlled-not (CCNOT) gate on three 
qubits:
\begin {eqnarray}
U_{\rm CCNOT}=\left(\begin{array}{cccccccc}
1& 0& 0& 0&0&0& 0& 0\\
0& 1& 0& 0&0&0& 0& 0\\
0& 0& 1& 0&0&0& 0& 0\\
0& 0& 0& 1&0&0& 0& 0\\
0& 0& 0& 0&1&0& 0& 0\\
0& 0& 0& 0&0&1& 0& 0\\
0& 0& 0& 0&0&0& 0& 1\\
0& 0& 0& 0&0&0& 1& 0\\
\end{array}\right).
\end{eqnarray}
This may be implemented with a sequence of five controlled rotations 
\cite{Barenco95}.  However using a line-selective pulse it is straightforward 
to 
generate 
the key part of this gate \cite{Linden98a}, giving the unitary operator
\begin {eqnarray}
\widetilde U_{\rm CCNOT}=\left(\begin{array}{cccccccc}
1& 0& 0& 0&0&0& 0& 0\\
0& 1& 0& 0&0&0& 0& 0\\
0& 0& 1& 0&0&0& 0& 0\\
0& 0& 0& 1&0&0& 0& 0\\
0& 0& 0& 0&1&0& 0& 0\\
0& 0& 0& 0&0&1& 0& 0\\
0& 0& 0& 0&0&0& 0& i\\
0& 0& 0& 0&0&0& i& 0\\
\end{array}\right).
\end{eqnarray} 
It is not difficult to show that 
\begin{eqnarray}
U_{\rm CCNOT} = e^{i{\pi\over 4}  I_z +i{\pi\over 4}   S_z - i{\pi\over 4} 2I_z 
S_z}
\ \widetilde U_{\rm CCNOT},
\end{eqnarray}
up to an irrelevant overall phase. Thus the exact unitary operation required, 
$U_{\rm 
CCNOT}$, can be implemented by performing $\widetilde U_{\rm CCNOT}$ followed 
by 
a 
sequence of corrections of the type we  describe in this letter.

\section{Experimental realization}

	We take as representative practical example a homonuclear 
weakly-coupled three-spin 
($ISR$) system where all three coupling constants $J_{IS},\ J_{IR}$
and $J_{RS}$ are non-zero and 
resolvable. Relaxation effects (spin-spin and spin-lattice relaxation and 
the nuclear 
Over-hauser effect) are specifically excluded from the discussion since they  
change 
intensities in an 
irreversible manner; of course if the pulse sequences have appreciable 
duration, 
relaxation effects will be observed.  Multiplet-selective radiofrequency 
refocusing pulses 
are employed in 
a symmetrical arrangement to refocus the appropriate chemical shifts and 
spin-spin 
splittings.  An essential property of these soft pulses is that they should 
act as ``universal 
rotors'', that is to say, within their effective bandwidth, they induce the 
same 
rotation whatever the initial 
state of the 
nuclear magnetization.   Gaussian shaped pulses do not satisfy this 
requirement; 
instead time-symmetric, shaped radiofrequency ``RE-BURP'' pulses 
\cite{Geen91} are used.  These 
refocusing ``$\pi$'' pulses are never applied to two different nuclear spins 
simultaneously, since 
that would introduce spin-echo modulation through the scalar coupling and 
create multiple-quantum coherence  through the double-resonance two-spin effect 
(TSETSE) 
\cite{Kupce94}.  The 
refocusing pulses are employed in a sequence designed to take advantage of 
the fact that 
radiofrequency pulse imperfections are corrected on even-numbered spin 
echoes \cite{Levitt81a}.   

Experiments were carried out at 22šC on a Varian 400 MHz 
high-resolution 
spectrometer on the three coupled protons ($R,\ I$ and $S$) in 
2,3-dibromopropanoic acid 
dissolved in benzene-${\rm d}_6$, where $J_{IS}=-10.1$ Hz, $J_{IR}=11.3$ Hz,
$J_{RS}=4.3$ Hz,  $\delta_{IR} = 188.5$ Hz 
and $\delta_{IS} = 219.5$ Hz. 
After removal of dissolved oxygen by bubbling argon through the 
solution, the measured spin-lattice relaxation times were 9.5 s ($R$) and 
1.8 s ($I$ and $S$).  
Although this particular homonuclear system provides a simple example to 
demonstrate the 
principles involved, it is not ideal for the purpose.   The chemical shift 
differences
are rather small, 
placing heavy demands on the selectivity of the soft $\pi$ pulses, the 
uniformity 
of excitation across 
a chosen spin multiplet, and the effectiveness of the suppression 
off-resonance.   Some 
strong coupling effects are also evident, particularly between spins $R$ and 
$I$.   Analogous 
experiments on heteronuclear systems (or on homonuclear 
carbon-13 
systems) would be expected to pose fewer practical problems.  Selectivity 
would no longer 
be crucial, and because the pulses would be of much shorter duration, 
relaxation and 
coupling effects during the pulses could be safely neglected.

\subsection{Module {\bf O1}: Selective phase evolution }
The basic module {\bf O1} is designed to refocus the spin-spin splittings 
of the active ($I$) 
spin but allow chemical shift evolution to accumulate a phase shift $\phi$, 
leaving the coupled 
spectators $R$ and $S$ unaffected (Fig. 1).  After excitation of magnetization 
by a hard pulse applied about the $x$-axis, the 
divergence of $I$-spin magnetization trajectories due to $J_{IR}$ and 
$J_{IS}$  is refocused by soft  $\pi$
pulses applied to $R$ and $S$, so that only the chemical shift effect 
persists.  It was found that 
the effects of pulse imperfections were reduced by applying the two $S$ 
pulses with the same 
phase ($+y$) and by setting the phases of the four $R$ pulses to $(+y\  +y\ -y\ 
-y)$ 
according 
to the MLEV-4 
prescription \cite{Levitt81b}.  The phase evolution diagram (Fig. 1) 
illustrates the refocusing process. The 
$S$ and the $R$ 
spins remain passive spectators because their chemical shifts and spin-spin 
splittings are 
refocused by the $\pi$ pulses, and certain pulse imperfections are compensated 
by the operation of 
the $\pi$ pulses acting in pairs \cite{Levitt81a}.   At the end of the 
sequence the $I$ spins accumulate a phase 
shift $\phi=2\pi\delta_I 8\tau$ radians which can be controlled by adjusting 
either the duration of the 
sequence $(8\tau)$ or the offset $\delta_I$ Hz between the $I$-spin chemical 
shift and the transmitter 
frequency.  

	There is a minor instrumental complication when soft radiofrequency 
pulses are used 
in this manner, since each radiofrequency field induces a small 
Bloch-Siegert shift \cite{Bloch40} on 
neighbouring resonances, always in the direction to move the neighbour 
resonance away 
from the irradiation field.  This translates into a small additional 
increment in the phase of 
the nuclear precession.   For the $I$ spins these small phase errors merely 
accumulate and 
can be corrected by 
a slight readjustment of the overall duration of the sequence.  For the $S$ 
spins the symmetry 
of the sequence is such that these phase errors are effectively refocused by 
the soft $\pi$ pulses.   
Unfortunately the perturbation of the motion of the $R$ spins by the pulses 
applied to $S$ is not 
refocused.   The Bloch-Siegert effect on $R$ was therefore compensated by 
applying ``ghost''  $\pi$
pulses at the same time as the $S$ pulses and with the same amplitude and 
duration, but with 
the offset from the $R$ resonance reversed in sign.   These ghost pulses 
fall in empty regions 
of the spectrum, exciting no NMR response directly.  With this precaution an 
experimental 
test of the $e^{i\phi I_z}$ module gave the spectra illustrated in Fig. 3. The 
duration of the sequence 
$(8\tau)$ was incremented so that the accumulated phase $\phi$ of the 
$I$-spin response varied over a range of $2\pi$ radians.   Note the attenuation 
of the 
$I$-spin signal due to $T_2^*$ losses over 
the relatively long durations ($8\tau$ varied between 1.15 and 1.23 
seconds), whereas this effect 
is refocused for the $R$ and $S$ responses.   

	A ``do-nothing'' module  that refocuses all chemical 
shift evolutions and 
spin-spin splittings can be fashioned from the module which creates $e^{i\phi 
I_z}$ by choosing 
$8\tau\delta_I$ to be an integral number 
of complete rotations, (as is essentially the case in Fig 3(a) and 3(e)).  A 
{\em general} ``do nothing'' sequence, valid for any duration $8\tau$, can be 
constructed 
by introducing 
two hard (non-selective) $\pi$ pulses at times $4\tau$ and $8\tau$, where 
the trajectories of $I,\ R$ and $S$ are 
at a focus (Fig. 1).

\subsection{Module {\bf O2}: Selective rotation }
	The {\bf O1} phase evolution sequence (a rotation about the $z$ axis) 
can be converted into a 
module which produces rotation of a given spin by an arbitrary angle about an 
arbitrary axis (with no effect on the other spins). The principle of 
``decoupling'' the rotation of the active ($I$) 
spins 
from the motion of 
the spectator spins ($R$ and $S$) was tested on the protons of 
2,3-dibromopropanoic acid by 
imposing an $I$-spin rotation through $\phi=\pi$ radians about an axis 
in the $xz$ plane tilted at $\beta=\pi/4$ radians.  
This takes $+z$ magnetization to the $+x$ axis, for example.  These 
experiments exploited an improved RE-BURP pulse (effective bandwidth 25 Hz), 
re-optimized to give more uniform re-focusing and better out-of-band 
suppression.  Precautions were 
 taken to 
compensate the effects of spin-lattice relaxation (which acts unequally 
during the different $\tau$ 
intervals) by increasing the tilt angle $\beta$ beyond $\pi/4$ radians. The 
experimental results (Fig. 4a) confirm the principle of decoupled rotation, 
showing a strong excitation of the $I$-spin 
response and residual responses from the spectators $R$ and $S$ reduced by 
more than an order 
of magnitude.  Simulations were performed for this sequence based on 
numerical 
integration of the Liouville-von Neumann equation, neglecting relaxation but 
allowing for 
strong coupling.  They indicate similar levels of residual responses for $R$ 
and $S$ (Fig. 4b).

\subsection{Module {\bf O3}: Selective evolution under the scalar coupling}
The sequence shown in Fig. 2 
implements the 
principal part of a CNOT gate, namely the unitary operator
$
e^{i\phi 2I_z S_z}.
$
Hard $\pi$ pulses are applied at times 
$2\tau$ and $6\tau$, where the $J_{IR}$ 
splittings of the $I$-spin response and the $J_{SR}$ splitting of the 
$S$-spin response are at a focus.  In this experiment the preparation stage 
involved the excitation of $y$ magnetization.
The $I$-spin and $S$-spin chemical shifts are refocused but the splitting 
$J_{IS}$ continues to generate 
diverging trajectories throughout the sequence, giving rise to a phase 
evolution of $\pm 8\pi\tau J_{IS}$ 
radians on both the $I$-spin and $S$-spin responses.  This is illustrated in 
Fig. 5 for phase 
increments of $\pm \pi/2$ radians.   On traces (b) and (d) the (antiphase) 
$I$ and $S$ responses would 
have appeared in dispersion; for the sake of clarity the receiver phase was 
changed by $\pi/2$ to 
reset them to the absorption mode.    The $R$ spin response remains 
essentially unchanged 
throughout.

\section{``Much Ado About Nothing?''}

        Whereas classic NMR experiments commonly ignore collateral
perturbation of coupled neighbour spins, a quantum computer requires that
all these ``spectator'' spins return to their initial states at the end of
the manipulation.   For pulse sequences that are not instantaneous, this
 ``do nothing'' operation is by no means trivial; it necessitates exact
refocusing of chemical shifts and spin-spin splittings.

We have
demonstrated how selectivity can be achieved in a homonuclear coupled 
three-spin
system for three important basic operations -- phase evolution, an
arbitrary rotation, and evolution under the scalar coupling.  Note
that these operations do not assume that the spin system is initially at
Boltzmann equilibrium; the new ``modules'' are unitary
operators that act on an arbitrary initial state of the spin system.

For NMR to be a useful technology for quantum computation, it will be necessary 
to 
extend current ideas to systems of many more spins.  Indeed recent work  
suggests 
that only for such extended systems will the computations be fully quantum 
mechanical  \cite{Caves98}.      The extension of these soft pulse 
sequences to accommodate more
spectator spins is straightforward, although in practice this lengthens the
duration of the sequence.   We must remember, however, that future
implementations of NMR quantum computers need not be restricted to groups
of coupled protons, so the selectivity requirements could be far less
severe than in the present proton experiments, implying much shorter
radiofrequency pulses.

\vfill\eject
\noindent{\large\bf Acknowledgments}

The authors are indebted to Dr Sandu Popescu for many illuminating discussions 
and 
to Dr $\overline{\rm E}$riks Kup\v ce for the design of the 
improved soft $\pi$ pulse (RE-BURP-Q).

} 

\vfill \eject
{\openup 5pt
\begin{center}
Figure Captions
\end{center}
Fig. 1. (Top) The {\bf O1} module.  This sequence of soft $\pi$ pulses 
(ellipses) is
designed to allow phase evolution of the four $I$-spin magnetization
components due to their chemical shift (dotted line) without spin-spin
splittings, returning the coupled $R$ and $S$ spins to their initial states.
(Bottom) Phase evolution diagram for the four $I$  lines.

Fig. 2.  (Top) The {\bf O3} module designed to focus the $I,\ S$ and $R$
 chemical
shifts but allow divergence of $I$ and $S$ magnetization components due to 
$J_{IS}$.
Ellipses represent soft pulses.  (Bottom)   The phase evolution of the
four $I$ lines.  

Fig. 3. Experimental test of the {\bf O1} module which allows phase 
evolution of
transverse magnetization of the $I$ spins without affecting the $R$ and $S$
responses.  The quiet version of the soft $\pi$ pulses was used (RE-BURP-Q)
with a bandwidth of 40 Hz.  The duration of the sequence (8$\tau$) was
incremented to give phase steps of $\pi/2$ radians over a range of $2\pi$.

Fig. 4.(a) Experimental test of an {\bf O2} module which rotates the $I$
magnetization through $\pi$ radians about an axis tilted through $\pi/4$ 
radians
in the $xz$ plane, leaving the $S$ and $R$ spins largely unaffected.  The 
RE-BURP-Q soft pulse had a bandwidth of 25 Hz
and a duration of 220 ms.  The soft pulse sequence of Fig. 1 was used, but
with the assignment of soft pulses to $S$ and $R$ interchanged.  In practice
the tilt angle was made appreciably larger than $\pi/4$ to compensate for
spin-lattice relaxation effects.  (b) Simulated spectrum
allowing for strong coupling but neglecting relaxation.

Fig. 5. Experimental test of the {\bf O3} module in which $J_{IS}$ causes phase 
divergence 
of
the $I$ and $S$ transverse magnetization spanning a range of  $\pm 2\pi$ 
radians in steps of $\pm \pi/2$ radians.  RE-BURP 
pulses with a 40 Hz bandwidth and a duration of 121 ms were used. In traces
(b) and (d) the receiver phase was  shifted by $\pi/2$
 for the $I$ and $S$
responses in order to reset them to pure absorption.

} 
\end{document}